\documentclass[prb,twocolumn,aps,showpacs,fixfloats]{revtex4}
\usepackage{graphicx}
\usepackage{bm}
\usepackage{amsmath,amssymb}
\usepackage{subfigure}
\usepackage{float}
\usepackage{latexsym}
\usepackage{color}
\usepackage{enumerate}
\usepackage{pdfpages}
\usepackage{tikz}
\usepackage{hyperref}
\usepackage{relsize}

\begin{document}
\newcommand{\s}{\scriptscriptstyle}
\newcommand{\uu}{\uparrow \uparrow}
\newcommand{\ud}{\uparrow \downarrow}
\newcommand{\du}{\downarrow \uparrow}
\newcommand{\dd}{\downarrow \downarrow}
\newcommand{\ket}[1] { \left|{#1}\right> }
\newcommand{\bra}[1] { \left<{#1}\right| }
\newcommand{\bracket}[2] {\left< \left. {#1} \right| {#2} \right>}
\newcommand{\vc}[1] {\ensuremath {\bm {#1}}}
\newcommand{\tr}{\text{Tr}}
\newcommand{\Trans}{\ensuremath \Upsilon}
\newcommand{\Refl}{\ensuremath \mathcal{R}}

\title{Landau-Zener transition in a two-level system coupled to a single highly-excited oscillator }

\author{Rajesh K. Malla, and M. E. Raikh}

\affiliation{ Department of Physics and
Astronomy, University of Utah, Salt Lake City, UT 84112}

\begin{abstract}
Two-level system strongly coupled to a single resonator mode (harmonic oscillator) is a paradigmatic model in many subfields of physics. We study theoretically the Landau-Zener transition in this model.
Analytical solution for the transition probability is possible
when the oscillator is highly excited, i.e. at high temperatures.
Then the relative change of the excitation level of the oscillator
in the course of the transition is small.
The physical picture of the transition in the presence of coupling
to the oscillator becomes transparent in the limiting cases of slow
and fast oscillator. Slow oscillator effectively renormalizes
the drive velocity. As a result,
the transition probability either increases
or decreases depending on the oscillator phase.
The net effect is, however, the suppression of the transition probability.
On the contrary, fast oscillator renormalizes the matrix element of the
transition rather than the drive velocity. This
renormalization makes the transition probability a non-monotonic function of the coupling amplitude.

\end{abstract}

\maketitle

\section{Introduction}
Since the publication of seminal papers
\cite{Caldeira,Chakravarty}, see also the
review Ref. \onlinecite{Review},
the effect of environment on the dynamics
of a two-level system
is modeled by introducing the coupling of the
levels to the infinite set of
harmonic oscillators.

In the course of the Landau-Zener (LZ) transition,\cite{Landau1932,Zener1932,Majorana,Stueckelberg}
when the energy levels of the two-level system undergo
the avoided crossing under the action
of external drive, the effect of environment amounts
to the loss of adiabaticity of the transition.
More quantitatively, the probability for the
system to stay in the ground state after the transition is
diminished
by the environment.\cite{Gefen,Ao1991,Shimshoni,Nalbach2009+,Pekola2011+,Ziman2011+,Vavilov2014+,Nalbach2014+,Nalbach2015+,Ogawa2017+,Arceci2017,we}
The underlying reason for this
is the absorption of ``quanta" of the environment.
This absorption leads to decoherence,
which suppresses the interference of different virtual
tunneling pathways.

The situation is more delicate when the  environment is represented by a single
oscillator.\cite{Wubs2005++,Wubs2006++,WubsEPL,Wubs2007++,Zueco,Ashhab1,Ashhab2,Sinitsyn,Sun}
In experiment, the role of such an oscillator, which is coupled to the two-level system,
is played e.g. by the transmission line resonator\cite{Experiment1}
or by the optical resonator.\cite{Experiment2}

In the absence of coupling, the
amplitude of the LZ transition can be viewed
as coherent superposition of many amplitudes
corresponding to virtual trajectories.
With coupling, each virtual transition is
accompanied by the
excitation of the oscillator.
On the other hand, the stronger oscillator
is excited,
the stronger is the feedback that it
exercises on the two-level system. Then it is a compound object,
two-level system dressed by many oscillator quanta,
that undergoes the LZ transition.

For the two-level system coupled to the environment with a continuous
spectrum only the weak-coupling regime is of interest. This is because,
upon increasing coupling, the interference is completely suppressed, so that
the transition probability, $P_{\s LZ}$, assumes the value  $P_{\s LZ}=1/2$.
On the contrary, when the two level system is coupled to a single oscillator, there is a wide domain of parameters when the coupling is strong
while $P_{\s LZ}$ is still a strong function of the drive velocity.

Nontriviality of the Landau-Zener (LZ) transition in the
presence of coupling to the oscillator is highlighted
by the exact result reported  in Ref. \onlinecite{Wubs2006++}.
This result pertains strictly to zero temperature when at
time $t\rightarrow -\infty$ the oscillators is
in the ground state. It was demonstrated in Ref. \onlinecite{Wubs2006++} that if the
two-level system starts in the state $\uparrow$ and
ends in the state $\uparrow$, then the oscillator remains
in the ground state at $t\rightarrow \infty$.
This result can be viewed as a manifestation of the ``no-go"
theorem\cite{Sinitsyn2004,Shytov2004,Vitanov2005} in application
to the spin-boson model. Certainly, at intermediate times, the oscillator
can be excited. As a consequence of the above restriction, the
two-level system and the oscillator end up entangled.

Another manifestation of nontriviality of the LZ transition with coupling to
a single oscillator is the dependence of  $P_{\s LZ}$ on the coupling strength, $g$. In particular, numerical results of Ref. \onlinecite{Ashhab1} suggest that, for finite-temperature oscillator, the dependence of $P_{\s LZ}$ on $g$
is a non-monotonic curve with a minimum. In other words, upon increasing $g$,
the adiabaticity of the
transition first decreases and then increases again.
There is no clear physical picture explaining the emergence of this minimum.
In theory, coupling to the oscillator turns a single avoided crossing,
taking place in the course of the LZ
transition, into a network of avoided crossings\cite{GaraninNetwork}
corresponding to different oscillator levels.
The coupling strength quantifies the ``talking"
between the  $\uparrow$ and $\downarrow$ amplitudes pertaining to a certain oscillator level to the corresponding amplitudes for two neighboring oscillator levels.

In general, the problem of the LZ transition in the presence of coupling to the oscillator
contains, in addition to $g$, three other  parameters with the dimensionality of frequency:
the
matrix element between $\uparrow$ and $\downarrow$ levels, the
inverse bare LZ transition time, and the oscillator frequency.
Definitely, it is impossible to derive an analytical expression for
$P_{\s LZ}$ for arbitrary relations between these parameters.
In the present paper we focus on the situation when the oscillator
is highly excited. Under this simplifying assumption we identify
the domain of parameters where the asymptotic analytical expression
for $P_{\s LZ}$ can be found. Roughly speaking, the two domains
correspond to slow and fast oscillator depending on whether the
LZ transition time is shorter or longer than the oscillator period.
Coupling to a slow oscillator effectively renormalizes
the drive velocity.
In the case of a fast oscillator,
the LZ transition splits, as a results of coupling, into a sequence of individual
transitions even-spaced in time. The corresponding gaps
are the oscillating functions of coupling strength, $g$.
Non-monotonic behavior of the survival probability with $g$
is the result of interference of partial
transition amplitudes.
We confirm this behavior by solving the
many-level Schr{\"o}dinger equation numerically.

\begin{figure}
\includegraphics[scale=0.022]{timeplot_smallw.pdf}
\includegraphics[scale=0.050]{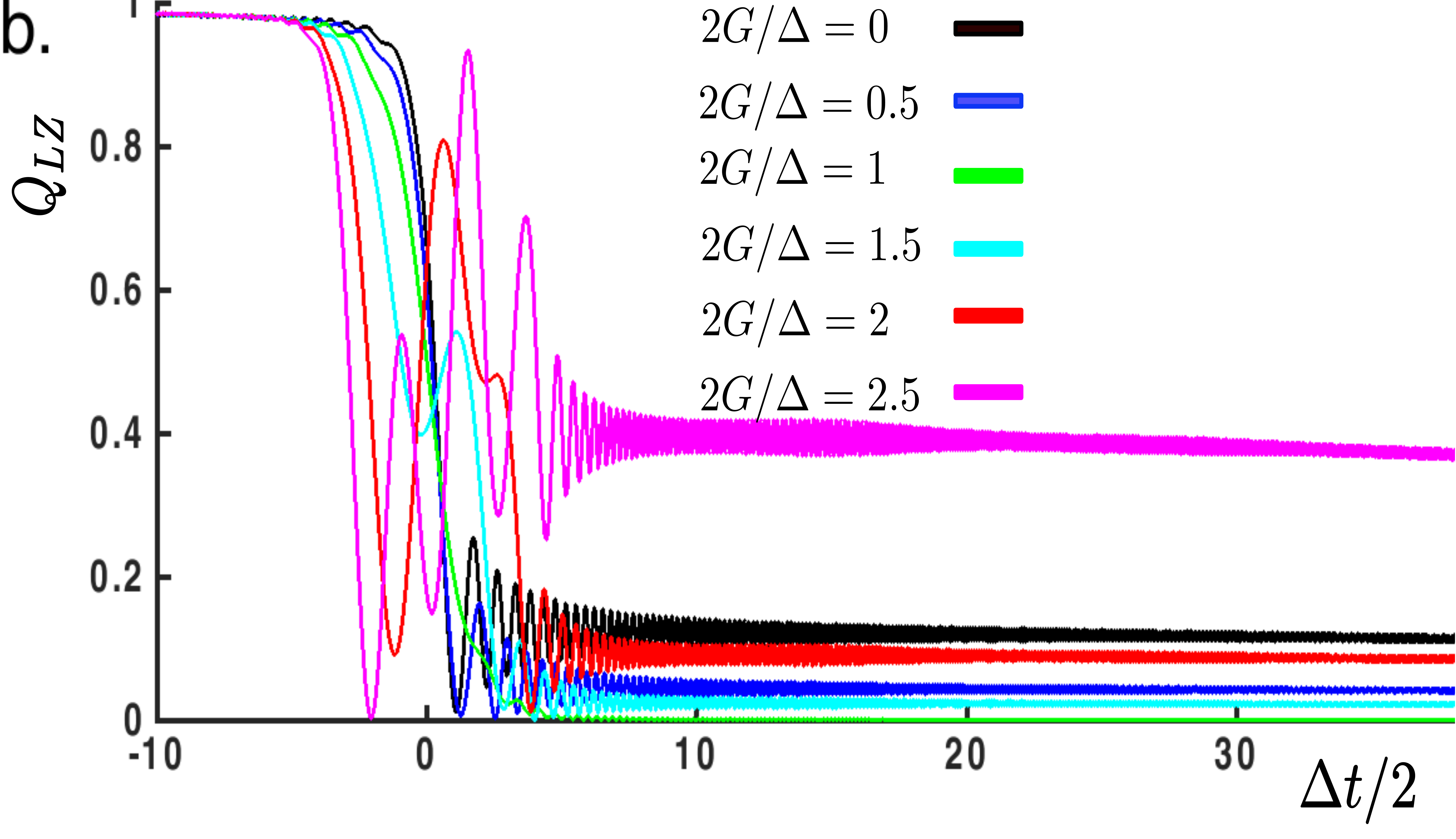}
\caption{(Color online) Survival probability, $Q_{\s LZ}$, calculated numerically from the
system Eq. (\ref{MainEqfinalform}) is plotted versus dimensionless time, $\Delta t/2$,
for several values of the coupling amplitude, $G$,
and for the oscillator phases
$\kappa =\pi/2$ (a) and $\kappa =-\pi/2$ (b). The oscillator frequency is chosen to be $\omega=0.25 \Delta$, where $\Delta$ is the gap at $t=0$,
while the drive velocity is chosen to be $v=\pi\Delta^2/4$. At zero coupling, the survival probability at $t\rightarrow \infty$ is $Q_{\s LZ}(\infty)=e^{-2}$. For $\kappa =\pi/2$ the values  $Q_{\s LZ}(\infty)$ grows monotonically  with $G$, while for  $\kappa =-\pi/2$  the value $Q_{\s LZ}(\infty)$ first drops with $G$ and then grows
with $G$.}
\label{timeplot}
\end{figure}

\begin{figure}
\includegraphics[scale=0.065]{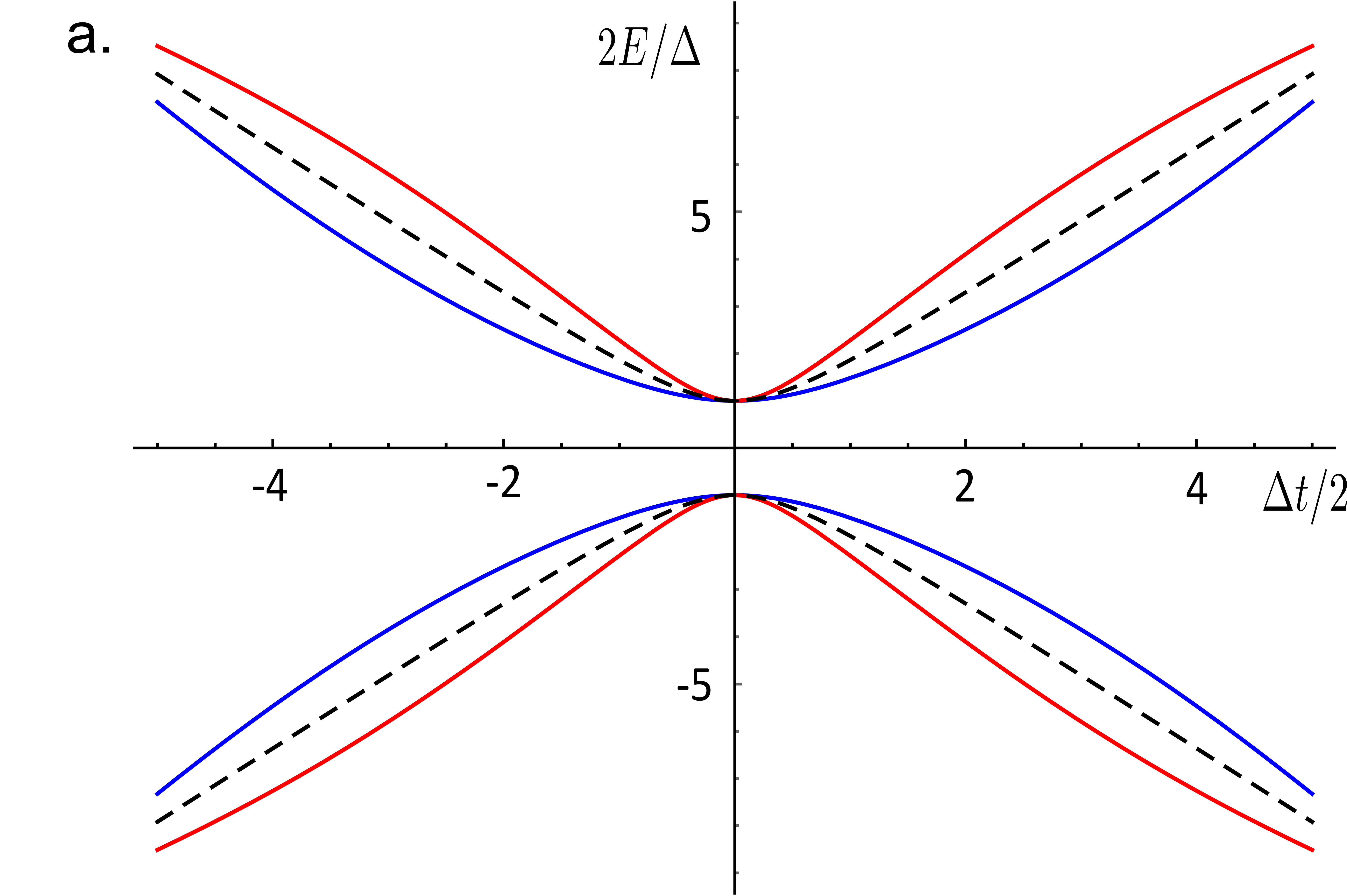}
\includegraphics[scale=0.04]{small_omega_action.pdf}
\includegraphics[scale=0.06]{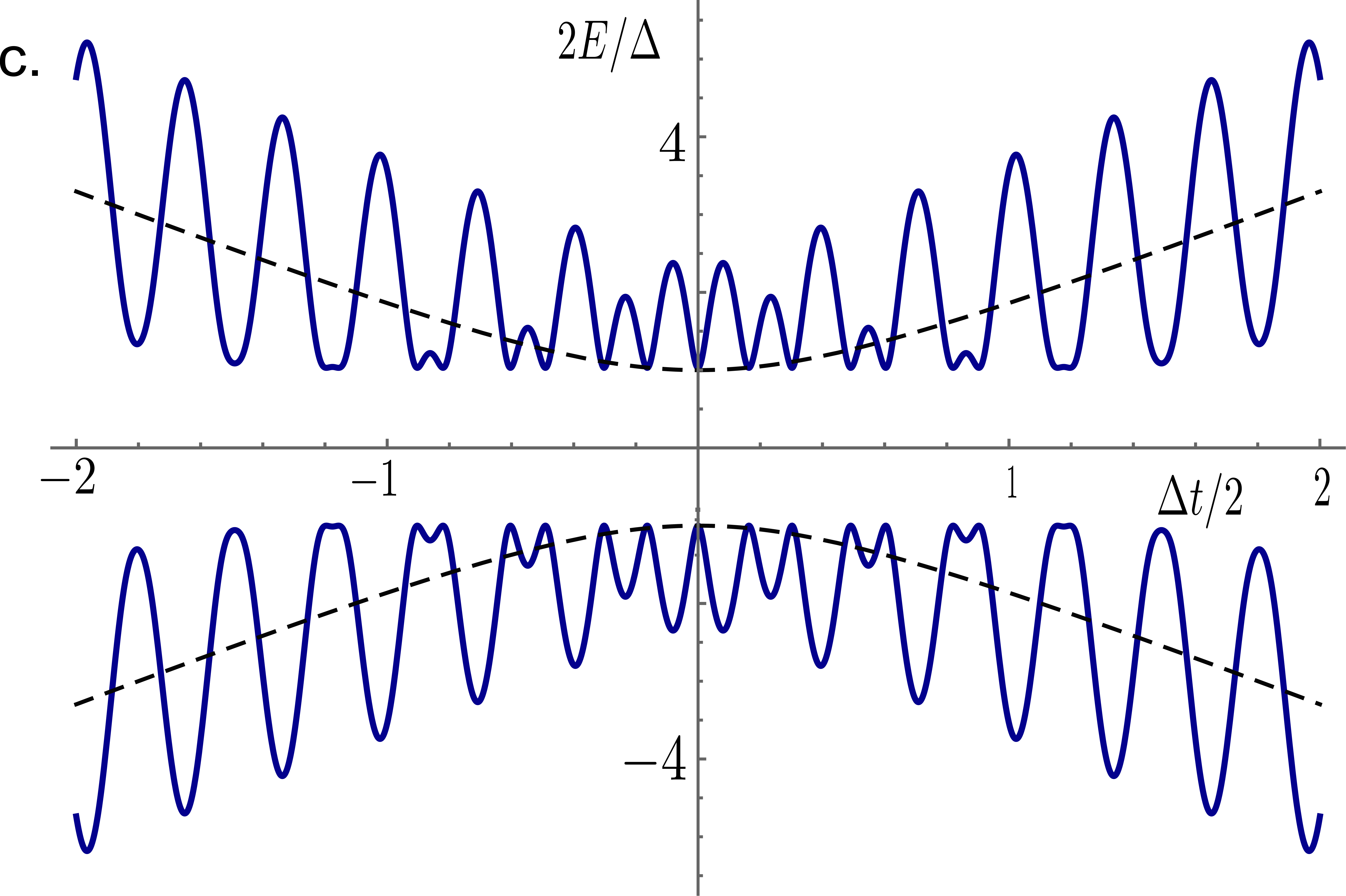}
\caption{(Color online)
The time-dependent energy levels Eq.~(\ref{E(t)}) are plotted
for $\kappa=\pi/2$ and different oscillator frequencies and the coupling strengths:
(a) $G=0.25\Delta$, $\omega=0.25 \Delta$; (b) $G=4\Delta$, $\omega=0.25 \Delta$;
(c) $G=0.5\Delta$, $\omega=10 \Delta$.
The level positions in the absence of coupling are shown with dashed lines.
Three distinct shapes of the curves $E(t)$
illustrate three different scenarios of how the coupling to the oscillator affects
the LZ transition. (a) and (b) correspond to the slow oscillator regime.
In (a), the coupling amounts to effective modification of the drive
velocity. Upon increasing the coupling, (b), the LZ transition
is the result of interference of many isolated LZ transitions.
For the fast oscillator, (c), the description of the LZ transition in terms
of time-dependent energy levels is inadequate. Blue lines in (a) show the energy levels
calculated for the same coupling and frequency as red lines but for $\kappa=-\pi/2$.  }

\label{action}
\end{figure}
\section{Basic relations}
We start from the standard Hamiltonian of the spin-boson model with drive
\begin{equation}
\label{Hamiltonian}
{\hat H}=-\frac{vt}{2}{\hat \sigma_z} -\frac{\Delta}{2}{\hat \sigma_x} + \omega {\hat b}^{\dagger} {\hat b} +g {\hat \sigma_z}\otimes({\hat b}+{\hat b}^{\dagger}),
\end{equation}
where $\omega$ is the oscillator frequency, while ${\hat b}$ and ${\hat b}^{\dagger}$
are, respectively, the annihilation and the creation operators of the oscillator.
The drive is characterized by the rate, $v$, of the change of the energies of the
$\uparrow$ and $\downarrow$ states coupled directly by the matrix element $\Delta/2$.
It is assumed that the coupling between the two-level system and the oscillator
is longitudinal.

Denote with $a_{\s 1}^n$ and $a_{\s 2}^n$ the amplitudes to find the system in the states $\uparrow$ and $\downarrow$ with $n$ quanta excited. As follows from Eq. (\ref{Hamiltonian}), these amplitudes satisfy  the following infinite system of coupled equations

\begin{multline}
i{\dot a}_{\s 1}^n+\frac{vt}{2}a_{\s 1}^n+\Big(n+\frac{1}{2}\Big)\omega a_{\s 1}^n-\frac{\Delta}{2} a_{\s 2}^n
\\
=-g\Big[(n+1)^{\s 1/2}a_{\s 1}^{n+1}+n^{\s 1/2}a_{\s 1}^{n-1}\Big],\nonumber
\end{multline}
\vspace{-5mm}
\begin{multline}
i{\dot a}_{\s 2}^n-\frac{vt}{2}a_{\s 2}^n+\Big(n+\frac{1}{2}\Big)\omega a_{\s 2}^n -\frac{\Delta}{2} a_{\s 1}^n
\\
=g\Big[(n+1)^{\s 1/2}a_{\s 2}^{n+1}+n^{\s 1/2}a_{\s 2}^{n-1}\Big].\label{Amplitudeeq}
\end{multline}
Our goal is to find the analytical solution of this system in the limit when the oscillator
is highly excited, so that the relevant $n$-values are big. In this limit, we can neglect the
difference between $(n+1)^{\s 1/2}$ and $n^{\s 1/2}$.
Denote the initial state of the oscillator with $n=n_0\gg 1$.
A crucial simplification is achieved if, in the course of the transition,
the excitation level of the oscillator changes relatively weakly,
i.e. by $m$ quanta with $m$ much smaller than $n_0$. This allows to eliminate the explicit
$n$-dependence from the system Eq. (\ref{Amplitudeeq}).
Upon introducing the new variables

 \begin{align}
a_{\s 1}^n(t)=b_{\s 1}^m(t)\exp \Big[i\omega t \Big(n_0+m+\frac{1}{2}\Big)\Big] ,\nonumber \\ a_{\s 2}^n(t)=b_{\s 2}^m(t)\exp \Big[i\omega t \Big(n_0+m+\frac{1}{2}\Big)\Big] ,\label{modifiedAmplitude}
\end{align}
 we get
 \begin{align}
i{\dot b}_{\s 1}^m+\frac{vt}{2}b_{\s 1}^m-\frac{\Delta}{2} b_{\s 2}^m
&=&-G \Big[b_{\s 1}^{m+1}e^{i\omega t} +  b_{\s 1}^{m-1}e^{-i\omega t} \Big],\nonumber\\
i{\dot b}_{\s 2}^m-\frac{vt}{2}b_{\s 2}^m-\frac{\Delta}{2} b_{\s 1}^m
&=&G \Big[b_{\s 1}^{m+1} e^{i\omega t} +  b_{\s 1}^{m-1}e^{-i\omega t} \Big]. \label{ModifiedAmplitudeeq}
\end{align}
  where
 \begin{equation}
 \label{G}
 G=gn_0^{\s 1/2}.
 \end{equation}
Obviously, the partial solutions of the system Eq. (\ref{ModifiedAmplitudeeq}) are the
plane waves
\begin{equation}
b_{\s 1}^m(t) =B_{\s 1}(t)e^{-i\kappa m}, ~~~b_{\s 2}^m(t) =B_{\s 2}(t)e^{-i\kappa m},
\end{equation}
where $\kappa$ is the wave vector. The amplitudes $B_{\s 1}$, $B_{\s 2}$ satisfy the system
\begin{align}
i{\dot B}_{\s 1}+\Big[\frac{vt}{2}+2G\cos(\omega t-\kappa)\Big] B_{\s 1}   -\frac{\Delta}{2}B_{\s 2}=0,\nonumber\\
i{\dot B}_{\s 2}-\Big[\frac{vt}{2}+2G\cos(\omega t-\kappa)\Big] B_{\s 2}    -\frac{\Delta}{2}B_{\s 1}=0. \label{MainEqfinalform}
\end{align}
This system describes the LZ transition within a {\em given} $\kappa$.
The form Eq. (\ref{MainEqfinalform}) suggests the interpretation of $\kappa$
as a phase of the classical oscillator.

Original system Eq. (\ref{Amplitudeeq}) describes the ``spreading" of the initial
state with $n=n_0$ over the states with $n=n_0+m$. The survival probability
$Q_{\s LZ}^{n\rightarrow (n+m)}$ is the probability for the system, which starts at $t\rightarrow -\infty$ from a single nonzero amplitude $a_{\s 1}^{n_0}$, to remain in {\em one of the states}  $a_{\s 1}^{n_0+m}$ at $t\rightarrow \infty$. After reducing the original system to the form
Eq.   (\ref{MainEqfinalform}) the amplitude $B_{\s 1}$ represents a combination
$\sum\limits_m b_{\s 1}^m\exp(-i\kappa m)$. The initial condition that at $t\rightarrow -\infty$
the $m$-dependence of amplitude $b_{\s 1}^m$ is $\delta_{m,0}$ suggests that the solutions of
Eq.  (\ref{MainEqfinalform}) corresponding to different $\kappa$ have the same absolute value
at $t\rightarrow -\infty$. This allows to express the net LZ survival probability via the
survival probabilities corresponding to all $\kappa$-values
\begin{equation}
\label{overkappa}
Q_{\s LZ}=1-P_{\s LZ}=\sum\limits_m Q_{\s LZ}^{n\rightarrow (n+m)}=\int\limits_{-\pi}^{\pi}\frac{d\kappa}{2\pi}Q_{\s LZ}(\kappa).
\end{equation}
In the remainder of the paper we study the dependence of $Q_{\s LZ}$ found from Eqs. (\ref{MainEqfinalform}), (\ref{overkappa})
in the limits of slow and fast oscillator.


\begin{figure}
\includegraphics[scale=0.095]{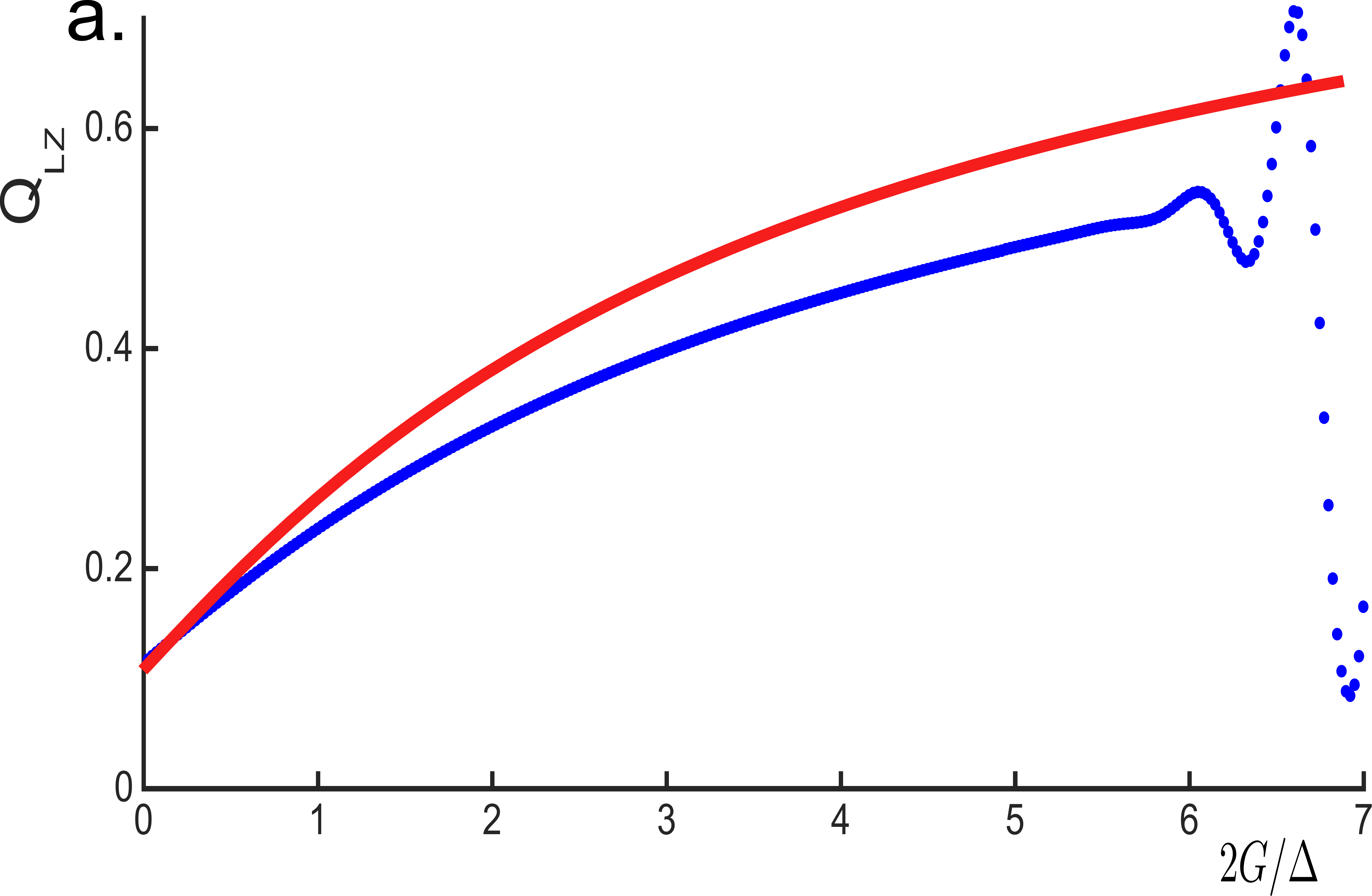}
\includegraphics[scale=0.095]{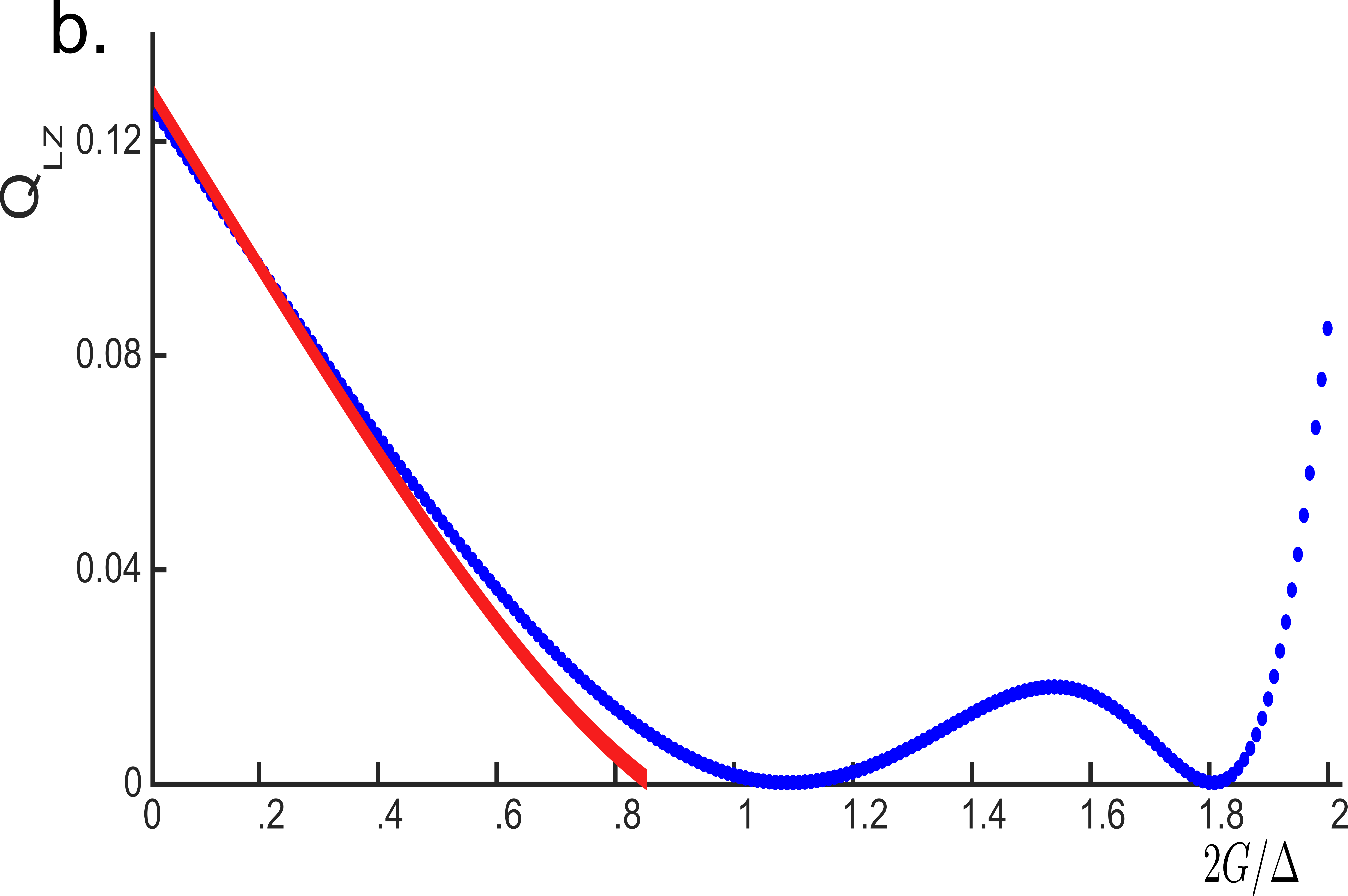}
\includegraphics[scale=0.095]{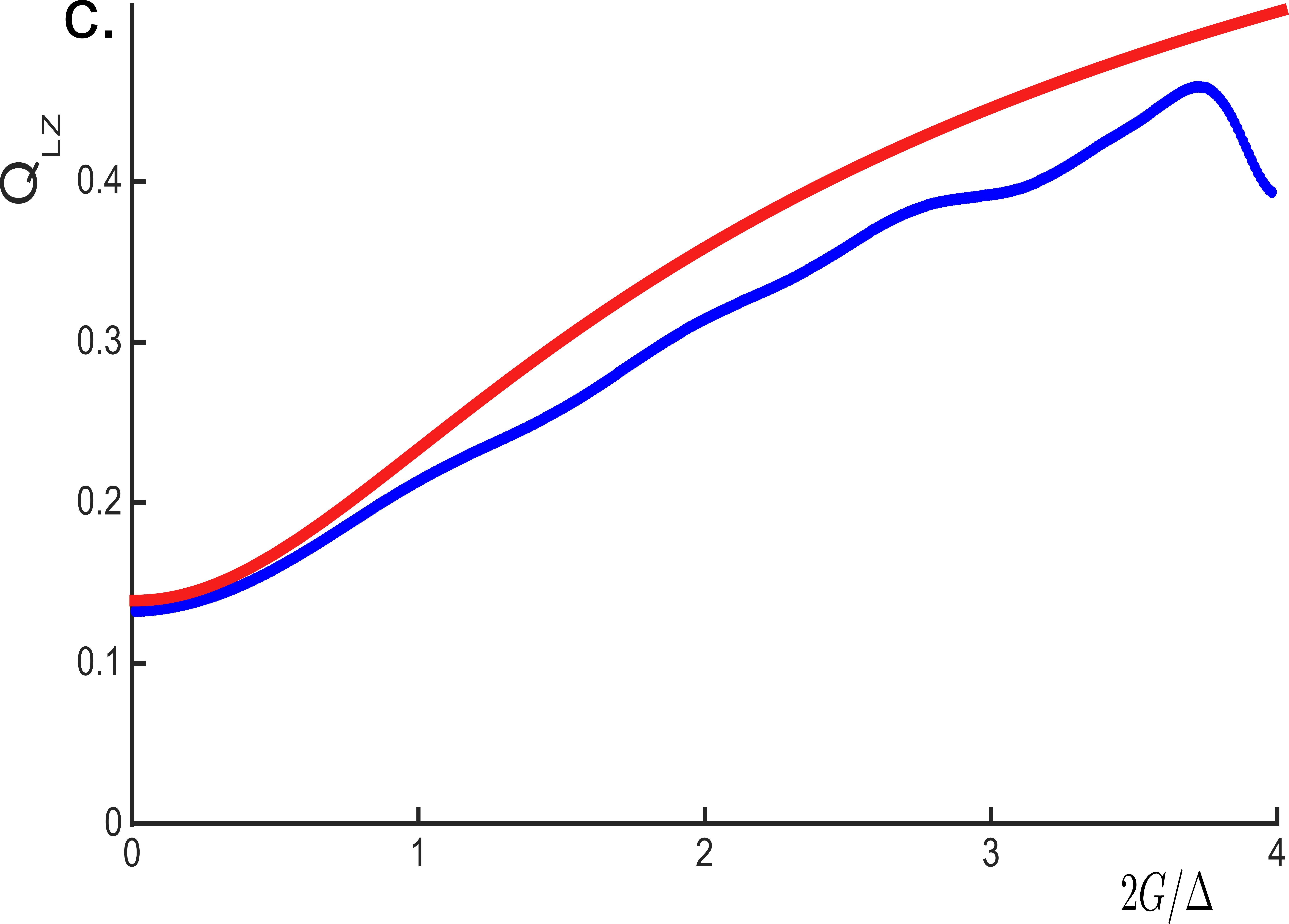}
\includegraphics[scale=0.12]{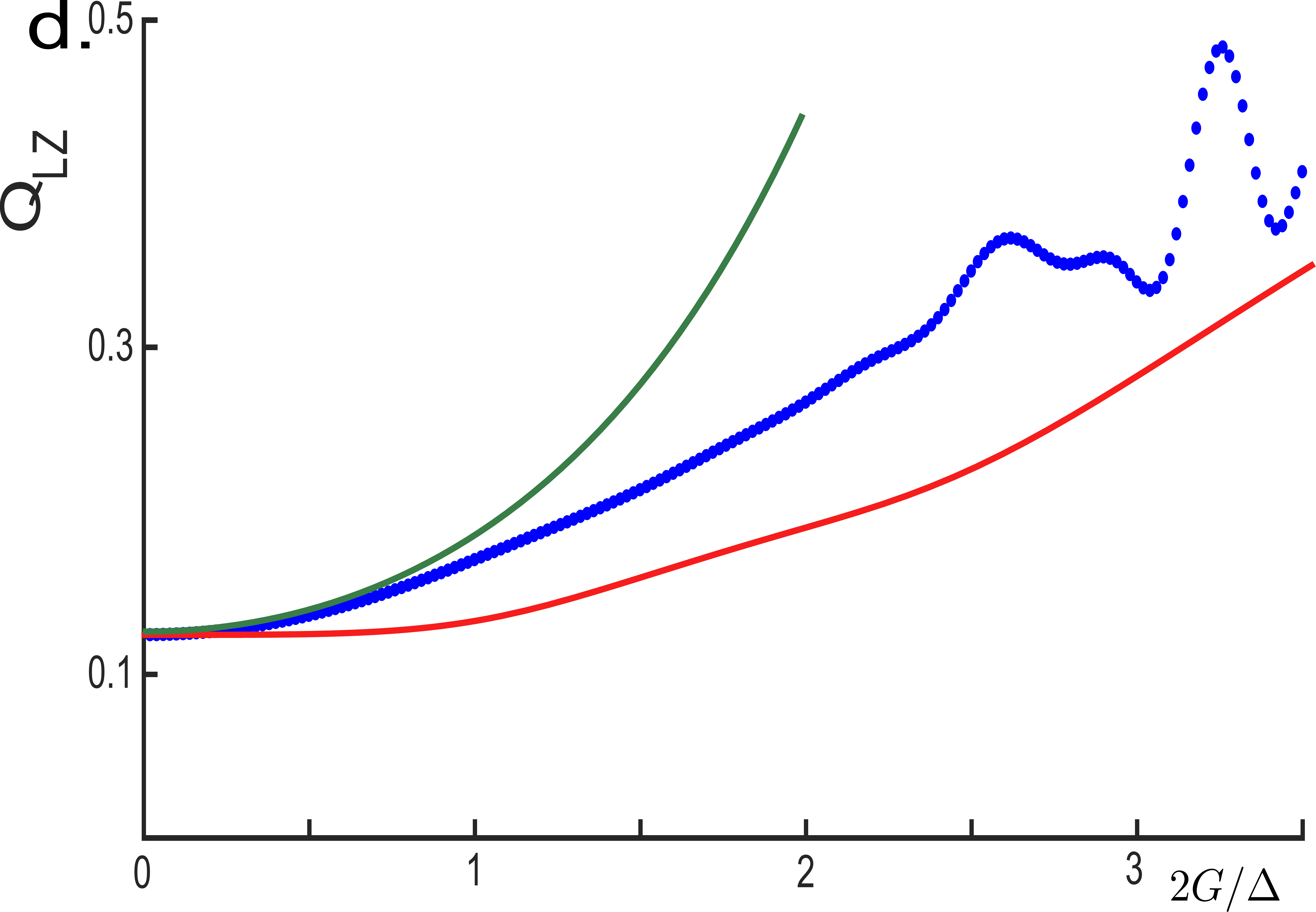}
\caption{(Color online) The survival probability, $Q_{\s LZ}$, at $t\rightarrow \infty$
is plotted versus the dimensionless coupling strength for the
 phases of oscillator $\kappa =\pi/2$ (a), $\kappa =-\pi/2$ (b), and $\kappa =0$ (c). Blue
 curves were obtained by solving the system Eq. (\ref{MainEqfinalform}) numerically.
Red curves are theoretical obtained from Eqs. (\ref{approximate}), (\ref{effective}). In the lower panel (d) the result of the numerical averaging of $Q_{\s LZ}(\kappa)$ is plotted with blue line. The green line shows the analytical result Eq. (\ref{approximate1}), while
the analytical result obtained from Eqs. (\ref{approximate}), (\ref{effective})
is shown with red line. The drive velocity and the oscillator frequency are the same as
in Fig. \ref{timeplot}.}
\label{smallomega}
\end{figure}

\section{Slow oscillator}

To illustrate how delicate is the effect of coupling to the low-frequency
oscillator on the survival probability, we plot in
Figs. \ref{timeplot}a, b
the numerical solutions of
the system Eq. (\ref{MainEqfinalform}) for different $G$-values. Fig. \ref{timeplot}a corresponds to the
wave vector $\kappa =\pi/2$,  while  Fig. \ref{timeplot}b corresponds to $\kappa =-\pi/2$.
The solutions $Q_{\s LZ}(t)$ correspond to
the ``subgap" frequency $\omega=0.25 \Delta$. The drive velocity is chosen to
be $v=\pi \Delta^2/4$, so that, without coupling, the survival probability $Q_{\s LZ}=\exp\left(-\pi\Delta^2/2v\right)$ is small,
$Q_{\s LZ}=e^{-2}$. Comparing Figs. \ref{timeplot}a and  \ref{timeplot} b, we conclude that the effect of
coupling on $Q_{\s LZ}$ is very different for these two values of $\kappa$. For $\kappa =\pi/2$
the survival probability increases monotonically with the coupling strength, $G$,
while for $\kappa =-\pi/2$, already for the minimal coupling
$G=0.25\Delta$, the value of $Q_{\s LZ}$ is smaller than in the absence of coupling.

Formal explanation of this peculiar dependence of $Q_{\s LZ}$ on $\kappa$ follows from the
expression of the time-dependent energy levels of the system Eq. (\ref{MainEqfinalform}).
This expression reads
\begin{equation}
\label{E(t)}
E=\pm \Bigg[\frac{\Delta^2}{4}+\left(\frac{vt}{2}+2G\cos\left(\omega t - \kappa \right)    \right)^2  \Bigg]^{1/2}.
\end{equation}
In Fig. \ref{action}a we plot these levels for
$G=0.25\Delta$ and $\kappa=\pm \pi/2$. It is seen that the plots $E(t)$ for
$\kappa =\pm \pi/2$ lie on the opposite sides of the $G=0$ curve. Thus, for $\kappa =\pi/2$, the coupling to the oscillator
effectively increases the drive velocity,
and, thus, $Q_{\s LZ}$ gets enhanced.
For $\kappa=-\pi/2$, the effective drive velocity
is decreased due to the coupling to the oscillator
and, correspondingly, $Q_{\s LZ}$ is diminished,


Dramatic difference of the survival probabilities for different
$\kappa$-values becomes even more dramatic
upon further increase of coupling strength.
This is illustrated in Fig. \ref{smallomega}, where the curves $Q_{\s LZ}(G)$
obtained numerically are plotted for $\kappa =\pi/2$, $\kappa =-\pi/2$,
and $\kappa =0$. While all three curves
start from $Q_{\s LZ}=e^{-2}$, the $\kappa =\pi/2$ and $\kappa =0$ curves
increase with $G$, while the $\kappa =-\pi/2$ curve decreases with $G$.
It also follows from Fig. \ref{smallomega} that beyond certain $G$-value all
three curves exhibit strong oscillations.

The goal of the theory is
to account for the shapes of the curves.
To this end, we recall that, in the absence of coupling,
the most concise way to derive expression
$Q_{\s LZ}=\exp\left(-\pi\Delta^2/2v\right)$ is to perform the analytic continuation of the semiclassical
solutions, $\exp\left(\pm i\int\limits_0^tdt'E(t') \right)$,  for the $\uparrow$, $\downarrow$ amplitudes
to the complex plane\cite{Dykhne1962}. Then  $Q_{\s LZ}$ emerges in the form of the
following integral between the turning points on the imaginary axis
\begin{equation}
\label{complexplane}
\ln Q_{\s LZ}=-\int\limits_{\tau_L}^{\tau_R}d\tau |E(i\tau)|=-\int\limits_{-\Delta/2v}^{\Delta/2v}d\tau\left[\frac{\Delta^2}{4}-\frac{v^2\tau^2}{4}\right]^{1/2}.
\end{equation}
Here $\tau_L$ and $\tau_R$ are the left and the right turning points, $E(i\tau_L)=E(i\tau_R)=0$,  which, in the absence of coupling,
are simply equal to $\pm \Delta/2v$.
General expression Eq. (\ref{complexplane}) suggests
that the extension to the finite coupling at
$\kappa =\pm\pi/2$ amounts to
the modification of
\begin{equation}
\label{E(tau)}
E(i\tau)\rightarrow \left[\frac{\Delta^2}{4}-\left(\frac{v\tau}{2}\pm 2G\sinh(\omega\tau)\right)^2\right]^{1/2}.
\end{equation}
The equation for $\tau_L$, $\tau_R$ becomes transcendental. Still, for a given set of parameters, the dependence
$Q_{\s LZ}(G)$ determined by Eqs. (\ref{complexplane}) and (\ref{E(tau)}) can be obtained by the numerical integration.
The results are shown in Fig.~\ref{smallomega}. They agree very well with $Q_{\s LZ}$ found from numerical solution of the system (\ref{MainEqfinalform}).

At weak coupling, the term $\pm 2G\sinh(\omega \tau)$ amounts to the modification of the drive velocity.
For the effective velocity obtained by expansion $\sinh(\omega\tau)$ at small $\tau$ we find
\begin{equation}
\label{effective}
v_{\s eff}=v\left(1\pm \frac{4G\omega}{v}\right).
\end{equation}
Upon substituting $v_{\s eff}$ into the LZ survival probability we get the results which agree perfectly with the result obtained
above using the semiclassical $E(i\tau)$, Eq. (\ref{complexplane}). This agreement could be expected only in ``perturbative" regime
$G\ll v/\omega$, but, for numerical reasons, this agreement holds up to the maximal value of $G=v/4\omega$. For this maximum value
$v_{\s eff}$ at $\kappa =-\pi/2$ turns to zero. Thus, we use this simplified procedure for arbitrary $\kappa$.
In doing this, it is very important to take into account that for $\kappa$ different from $\pm \pi/2$ the transition point is shifted from
$t=0$ to some $t=t_{\s eff}$.
The combination $\frac{vt}{2}+2G\cos(\omega t-\kappa)$ should be replaced by $\frac{1}{2}\left[v_{\s eff}(t-t_{\s eff})\right]$.
Fig. \ref{smallomega}c illustrates that theoretical dependence $Q_{\s LZ}(G)$ is in good agreement with numerical solution of the
system Eq. (\ref{MainEqfinalform}) at $\kappa =0$.

Summarizing, we write the expression for survival probability in the limit of a slow
oscillator in the form
\begin{equation}
\label{approximate}
Q_{\s LZ}(\kappa)=\exp\Bigg\{-\frac{\pi \Delta^2}{2v\left[1-\frac{4G\omega}{v}\sin(\omega t_{\s eff})    \right]}  \Bigg\},
\end{equation}
where $t_{\s eff}$ is determined by the condition
\begin{equation}
\label{effective}
\frac{vt_{\s eff}}{2}+2G\cos(\omega t_{\s eff}-\kappa)=0.
\end{equation}
The final step is averaging of Eq. (\ref{approximate}) over $\kappa$.
This averaging can be performed analytically when the renormalization
of the velocity due to the coupling to the oscillator is weak. Expanding
the denominator, we get

\begin{equation}
\label{approximate1}
Q_{\s LZ}=\exp\left(-\frac{\pi\Delta^2}{2v}\right)I_0\left(\frac{2\pi\Delta^2G\omega}{v^2}\right),
\end{equation}
where $I_0(z)$ is the modified Bessel function. We note that, while  deriving this result, we assumed that $G$ is much smaller than $v/\omega$, the argument of $I_0$ can be bigger than $1$. This is because
this argument contains an additional big factor $\pi\Delta^2/2v$.
In other words, Eq. (\ref{approximate1}) captures strong enhancement of the
survival probability caused by the coupling to the oscillator.

In Fig. \ref{smallomega}d we compare the results of three approaches to the
calculation of the  evolution of $Q_{\s LZ}$ with coupling strength.
The first result, shown with blue curve, is purely numerical. Namely,
the dependence $Q_{\s LZ}(\kappa)$ was obtained for each $G$-value
and then averaged over $\kappa$ numerically.
The second result (red line) is semi-analytical,
obtained from Eqs. (\ref{approximate}), (\ref{effective}), and,
finally, the analytical result Eq. (\ref{approximate1})
(green line). As could be expected, Eq. (\ref{approximate1}) captures
the  $Q_{\s LZ}(G)$ behavior only for small couplings.
The semi-analytical descriptions works well until $G \sim \Delta$.
The origin of the discrepancy between this description 
and the numerics is that $Q_{\s LZ}$ is the result of
the averaging of rapidly growing and rapidly decaying 
contributions.

In the limit of large $G\gtrsim v/\omega$ the curves in Fig. \ref{smallomega}
start to oscillate. The oscillations survive the averaging over $\kappa$.
The origin of these oscillations becomes clear from Fig.~\ref{action}b.
LZ transition at small $G$ evolves into a sequence of individual
well-resolved LZ transitions upon increasing $G$. The net number of transitions,
$N_s=4G\omega/\pi v$, grows linearly with coupling.
Passage of these transitions may result in constructive
or destructive interference depending on the
phase accumulated between the subsequent transitions.
The situation is fully analogous to the
Landau-Zener-St{\"u}ckelberg interferometry.\cite{ShevchenkoReports}

\section{Fast oscillator}


From Fig. \ref{action}c we realize that the description based
on time-dependent energy levels is inadequate for the fast oscillator. This is also clear from physical arguments, since the ``local" velocity is much bigger than the
drive velocity. The role of the oscillator
at large $\omega$ is to renormalize not the drive velocity but rather
the matrix element, $\Delta/2$, between the levels. To see this, we make the following substitution in the system Eq. (\ref{MainEqfinalform})

\begin{align}
\label{substitution}
B_{\s 1}(t)=D_{\s 1}(t)\exp\Bigg[\frac{2iG\sin(\omega t-\kappa)}{\omega}\Bigg],\nonumber \\
B_{\s 2}(t)=D_{\s 2}(t)\exp\Bigg[-\frac{2iG\sin(\omega t-\kappa)}{\omega}\Bigg],
\end{align}
after which it acquires the form

\begin{align}
\label{system1}
i{\dot D}_{\s 1}+\frac{vt}{2}D_{\s 1}=\frac{\Delta}{2}D_{\s 2}\exp\Bigg[-\frac{4iG\sin(\omega t-\kappa)}{\omega}\Bigg], \nonumber \\
i{\dot D}_{\s 1}-\frac{vt}{2}D_{\s 2}=\frac{\Delta}{2}D_{\s 1}\exp\Bigg[+\frac{4iG\sin(\omega t-\kappa)}{\omega}\Bigg].
\end{align}
In the form Eq. (\ref{system1}), the inter-level matrix element oscillates
with time. If the time of the LZ transition, $\sim \Delta/v$, is much longer
than the period of oscillations, the gap oscillates many times in the course of the
transition. This suggests
that the oscillating factor can be replaced\cite{Glenn} by its average $J_0(4G/\omega)$,
where $J_0$ is the Bessel function. This replacement immediately leads to
the survival probability
\begin{equation}
\label{Reduced}
Q_{\s LZ}^{(0)}=\exp\left[-\frac{\pi\Delta^2}{2v}
J_0^2\left(\frac{4G}{\omega}\right)\right].
\end{equation}
This result suggests that a very small $Q_{\s LZ}=e^{-2}$ at zero coupling
increases rapidly with coupling, reaches
$Q_{\s LZ}=1$, when the Bessel function passes through zero,
and then drops down.



\begin{figure}
\includegraphics[scale=0.11]{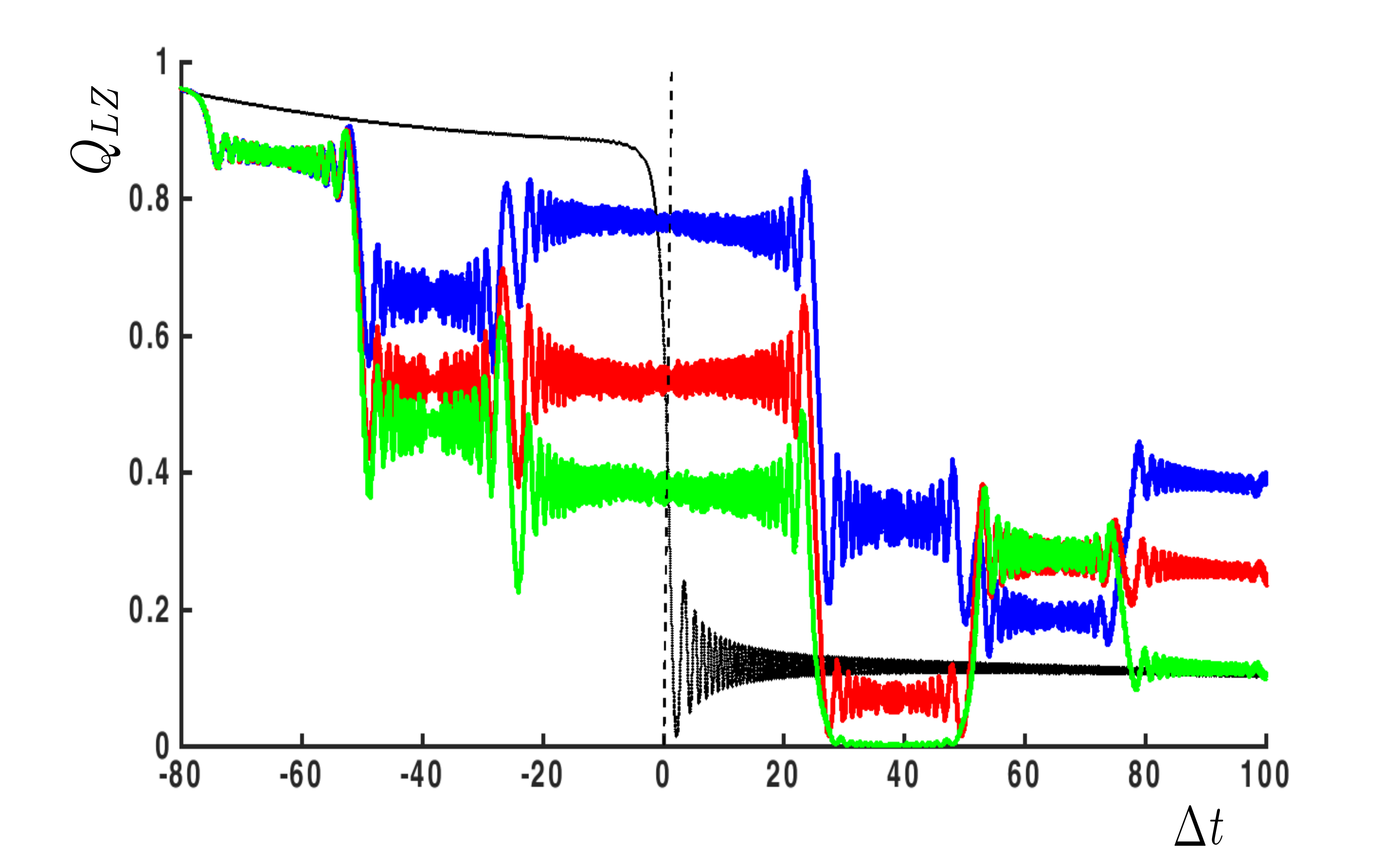}
\caption{(Color online) Survival probability, $Q_{\s LZ}$, calculated numerically from the
system Eq. (\ref{system1}), is plotted versus the dimensionless time, $\Delta t$,
for specific coupling strength $G=12\Delta$ and frequency $\omega=20\Delta$, so that
$\frac{4G}{\omega}=2.4$ and $J_0(4G/\omega)=0$.
Different curves correspond to the values of $\kappa$:
$\kappa =\pi/2$ (blue), $\kappa =\pi/3$ (red), and $\kappa =\pi/4$ (green).
It is seen that $Q_{\s LZ}$ does not change near $t=0$, where the $k=0$ LZ transition is expected suggesting that the
gap is suppressed. Black line shows the time evolution of $Q_{\s LZ}$ at zero coupling.
The drive velocity is $v=\pi \Delta^2/4$ in all the curves.}
\label{timeplotbessel}
\end{figure}

To check numerically the validity of averaging over
the oscillator period, in Fig. \ref{timeplotbessel} we show the time dependence  $Q_{\s LZ}(t)$
calculated by numerical solution of Eq. (\ref{system1}) for
particular value $\frac{G}{\omega}=0.6$, when the Bessel
function $J_0(4G/\omega)$ turns to zero. We see that, for this coupling,
$Q_{\s LZ}(t)$ does not change with time in a certain domain
around $t=0$ suggesting that the effective gap is zero.
However, unlike what Eq. (\ref{Reduced}) predicts,
the value of $Q_{\s LZ}$ is not one in this domain.
The reason is that the system Eq. (\ref{system1})
encodes a number of individual transitions
which take place around the times $t_k=k\omega/v$.
This becomes apparent if we make the following substitution
in the system Eq. (\ref{system1})
\begin{align}
\label{substitution}
D_{\s 1}(t)=e^{i\omega t/2}{\tilde D}_{\s 1}\left(t-\frac{\omega}{v}\right), \nonumber \\
D_{\s 2}(t)=e^{-i\omega t/2}{\tilde D}_{\s 2}\left(t-\frac{\omega}{v}\right).
\end{align}
Upon this substitution, Eq. (\ref{system1}) assumes the form

\begin{figure}
\includegraphics[scale=0.6]{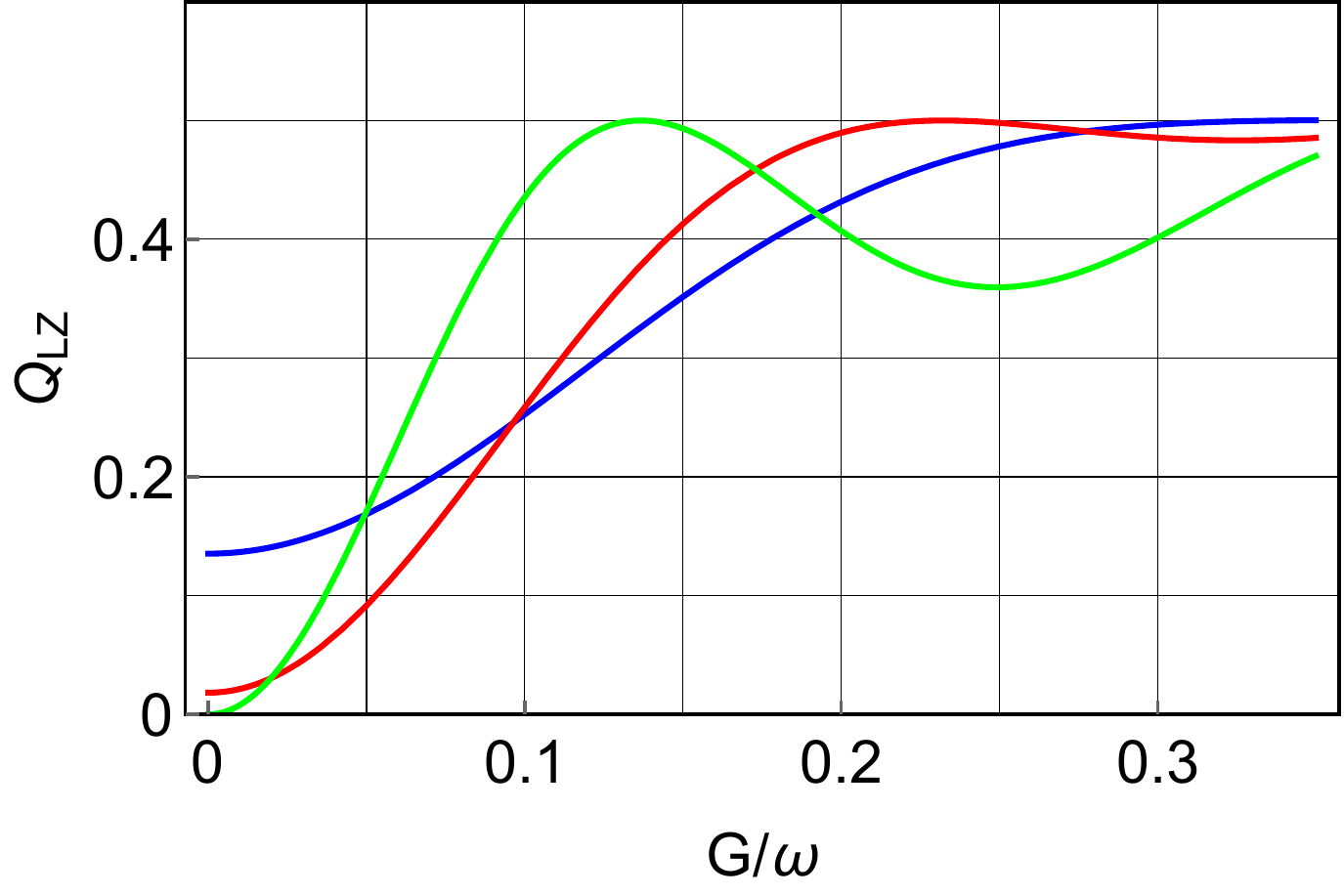}
\caption{(Color online) The ``incoherent" survival probability is plotted
from Eq. (\ref{Qlzincoherent}) versus the dimensionless coupling amplitude for different
values of the bare survival probability: $Q_{\s LZ}(0)=e^{-2}$ (blue),
 $Q_{\s LZ}(0)=e^{-4}$ (red), and $Q_{\s LZ}(0)=e^{-10}$ (green). For the
 latter curve the approach to the asymptote $Q_{\s LZ}(\infty)=\frac{1}{2}$
 is non-monotonic.}
\label{incoherent}
\end{figure}
\begin{figure}
\includegraphics[scale=0.66]{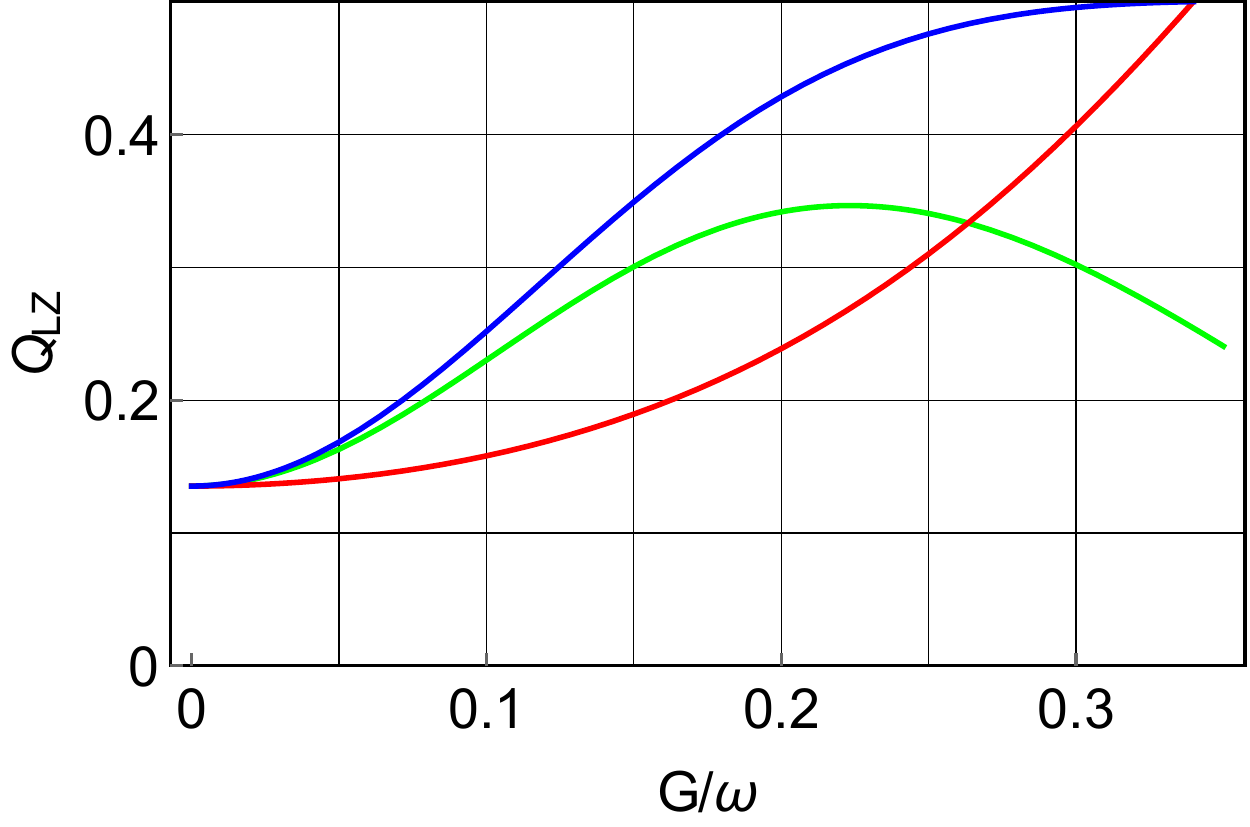}
\caption{(Color online)
Red: the survival probability, $Q_{\s LZ}^{(0)}$,  is plotted from
Eq. (\ref{Reduced}) versus the dimensionless coupling amplitude. The bare survival
probability is chosen to be $Q_{\s LZ}(0)=e^{-2}$;
Blue: the small-$G$ portion of the ``incoherent" result shown in Fig. \ref{incoherent};
Green: the dependence $Q_{\s LZ}^{(G)}$ with interference effects incorporated
is plotted from Eq. (\ref{pclassical1}).}
\label{classical}
\end{figure}

\begin{figure}
\includegraphics[scale=0.11]{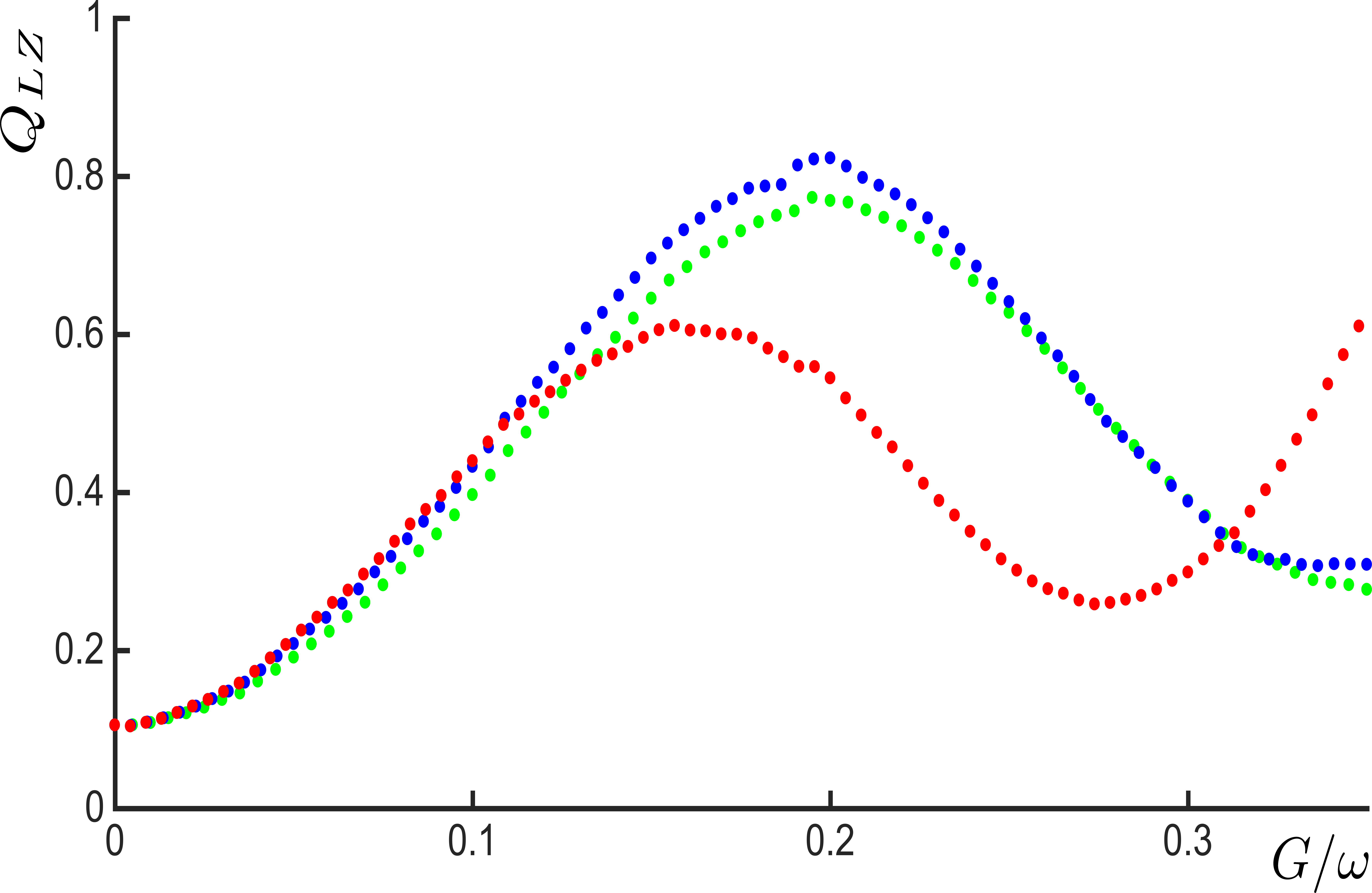}
\caption{(Color online) The net survival probability $Q_{\s LZ}(\infty)$ calculated by solving  numerically the system Eq. (\ref{system1}), and subsequently averaged over $\kappa$, is plotted versus the dimensionless coupling amplitude for three values of the oscillator frequency: $\omega=20\Delta$ (green), $\omega=22\Delta$ (blue), and $\omega=23\Delta$ (red). The bare $Q_{\s LZ}$ value is
chosen to be $Q_{\s LZ}=e^{-2}$, as in Fig. \ref{timeplotbessel}, which corresponds to
the drive velocity $v=\pi\Delta^2/4$. In the domain $\frac{G}{\omega}<0.1$ all three curves coincide and agree with the ``incoherent" and ``coherent" theoretical results shown in
Fig. \ref{classical}. The position of the maximum also agrees with ``coherent" curve
in Fig. \ref{classical}. However, the peak value $Q_{\s LZ}\approx 0.6$ is higher than
 $Q_{\s LZ}=0.35$ predicted by theory. Note that the theoretical result Eq.
 (\ref{pclassical1}) takes into account only $k=0$ and $k=\pm 1$ intermediate transitions.}
\label{bigomega}
\end{figure}

\begin{multline}
\label{system2}
i\dot{\tilde{D}}_{\s 1}\left(t-\frac{\omega}{v}\right)+
\frac{v}{2}\left(t-\frac{\omega}{v}\right)\tilde{D}_{\s 1}\left(t-\frac{\omega}{v}\right) \\
=\frac{\Delta}{2}\tilde{D}_{\s 2}\left(t-\frac{\omega}{v}\right)
\exp\Bigg[i\left(\omega t-\frac{4G\sin(\omega t-\kappa)}{\omega}\right)\Bigg],\nonumber
\end{multline}
\vspace{-5mm}
\begin{multline}
i\dot{\tilde{D}}_{\s 2}\left(t-\frac{\omega}{v}\right)-
\frac{v}{2}\left(t-\frac{\omega}{v}\right)\tilde{D}_{\s 2}\left(t-\frac{\omega}{v}\right) \\
=\frac{\Delta}{2}\tilde{D}_{\s 1}\left(t-\frac{\omega}{v}\right)
\exp\Bigg[-i\left(\omega t-\frac{4G\sin(\omega t-\kappa)}{\omega}\right)\Bigg].
\end{multline}
The same argument as above suggests that, as a result of being fast, the
exponent in the right-hand side can be  averaged over the period, $2\pi/\omega$,
Then the system Eq.~(\ref{system2}) will describe a regular LZ transition taking
place around $t=\frac{\omega}{v}$ with a gap reduced by $J_1(4G/\omega)$, where $J_1$
is the first-order Bessel function. The corresponding survival probability reads
\begin{equation}
\label{Reduced1}
Q_{\s LZ}^{(1)}=\exp\left[-\frac{\pi\Delta^2}{2v}
J_1^2\left(\frac{4G}{\omega}\right)\right].
\end{equation}
Naturally, a similar transition taking place around \newline $t=-\frac{\omega}{v}$
is described
by the same $Q_{\s LZ}^{(1)}$. Note also, that, in addition to $J_1(4G/\omega)$,
the averaged matrix element is multiplied by $\exp{(i\kappa)}$.
The physical meaning of the moments $t=\pm \frac{\omega}{v}$ is
transparent. The energy separation between $\uparrow$ and $\downarrow$ state
changes with time as $vt$. At $t=\frac{\omega}{v}$ this separation becomes
equal to the oscillator quantum.

The extension of Eqs. (\ref{Reduced}), (\ref{Reduced1}) to arbitrary $k$ is straightforward:
\begin{equation}
\label{Reduced2}
Q_{\s LZ}^{(k)}=\exp\left[-\frac{\pi\Delta^2}{2v}
J_k^2\left(\frac{4G}{\omega}\right)\right].
\end{equation}
The transitions taking place at $t=t_k$ can be viewed as well separated if the time, $\omega/v$, is much bigger than the individual
LZ transition time $\sim \Delta/v$, yielding the criterion $\omega \gg \Delta$.
The second condition to be met is that the effective time averaging
takes place during the
LZ transition time, so that $\Delta/v \gg 1/\omega$. The second condition can be cast in the form $\frac{\omega}{\Delta}\gg \frac{v}{\Delta^2}$. If the bare survival probability is small, $\Delta^2 \gg v$, then
the first condition is more restrictive.

Fig. \ref{timeplotbessel} offers an insight into a general scenario of the LZ transition
in a two-level system coupled to a fast oscillator. The system undergoes
a number of individual transitions at times $t=t_k$ characterized by survival probabilities $Q_{\s LZ}^{(k)}$. It is also seen from Fig. \ref{timeplotbessel} that the
evolution of $Q_{\s LZ}$ with time depends on the wave vector, $\kappa$.
This is a consequence of interference of partial transition amplitudes.
To find the net survival probability analytically, we first assume that
averaging over $\kappa$ suppresses the interference effects completely.
Then we can write the recurrent relation for $Q_N$, which are the
successive values of $Q_{\s LZ}$ after $N$ transitions

\begin{equation}
\label{recurrent}
Q_{N+1}=Q_N\left(1-Q_{\s LZ}^{(N+1)}\right)+\left(1-Q_N\right) Q_{\s LZ}^{(N+1)}.
\end{equation}
This relation expresses the fact that the occupation of the state $\uparrow$ after
the transition comes from the occupation of this state before the transition
as well as from the occupation of the state  $\downarrow$ which flips in the course
of the transition. It is straightforward to derive from Eq. (\ref{recurrent}) the expression for $Q_{\s LZ}$ after the arbitrary number of transitions
\begin{equation}
\label{Qlzincoherent}
Q_{\s LZ}=\frac{1}{2}-\frac{1}{2}\prod\limits_k \left(1-2Q_{\s LZ}^{(k)}\right).
\end{equation}
In Fig. \ref{incoherent} the behavior of $Q_{\s LZ}$ versus the coupling is plotted from
Eq. (\ref{Qlzincoherent}) for different bare values of $Q_{\s LZ}$.
Naturally, all the curves approach $1/2$ at very large coupling. This is because,
at large coupling, the information about the initial state of the two-level system
is erased. For the bare value $Q_{\s LZ}=e^{-2}$ the approach to $1/2$ is monotonic.
However, for $Q_{\s LZ}=e^{-4}$ a wiggle emerges at $\frac{G}{\omega} \approx 0.23$.
This feature evolves into a well pronounced maximum for bare $Q_{\s LZ}=e^{-10}$.

At the position of maximum in Fig. \ref{incoherent} (green line) the argument, $4G/\omega$, of the Bessel functions is about $0.6$. For this value and also for smaller couplings we can restrict
consideration to only three LZ transitions taking place at $t=0$ and $t=\pm \omega/v$,
since all higher Bessel functions are small. With only three transitions, we can incorporate  the interference effects into the theory. To do so, we should take into
account that each partial LZ transition is characterized by a scattering matrix which
contains the survival  probability and a phase, $\chi$. The evolution of the amplitudes to find
the system in $\uparrow$ and $\downarrow$ states is described by the product of scattering matrices

\begin{widetext}
\label{matrix}
\begin{equation}
M=\begin{pmatrix}
\sqrt{Q_{\s LZ}^{(1)}} & \hspace{-5mm}-i \sqrt{1-Q_{\s LZ}^{(1)}} e^{i \chi_{\s 3}} \\ \\ \\
-i \sqrt{1-Q_{\s LZ}^{(1)}} e^{-i \chi_{\s 3}} & \hspace{-5mm}\sqrt{Q_{\s LZ}^{(1)}}
\end{pmatrix}
\begin{pmatrix}
\sqrt{Q_{\s LZ}^{(0)}} & \hspace{-5mm}-i \sqrt{1-Q_{\s LZ}^{(0)}} e^{i \chi_{\s 2}} \\ \\ \\
-i \sqrt{1-Q_{\s LZ}^{(0)}} e^{-i \chi_{\s 2}} & \hspace{-5mm}\sqrt(Q_{\s LZ}^{(0)})
\end{pmatrix}
\begin{pmatrix}
\sqrt{Q_{\s LZ}^{(1)}} & \hspace{-5mm}-i \sqrt{1-Q_{\s LZ}^{(1)}} e^{i \chi_{\s 1}} \\ \\ \\
-i \sqrt{1-Q_{\s LZ}^{(1)}} e^{-i \chi_{\s 1}} & \hspace{-5mm}\sqrt{Q_{\s LZ}^{(1)}}
\end{pmatrix},
\end{equation}
\end{widetext}
From this product one can infer the following expression for the
amplitude to change the level after the three transitions\cite{Robert}


\begin{multline}
\label{A}
A_{\s \uparrow \rightarrow \downarrow} = -i Q_{\s LZ}^{(1)}\sqrt{1-Q_{\s LZ}^{(0)}} e^{-i \chi_{\s 2}} \\
 - i\sqrt{Q_{\s LZ}^{(1)}Q_{\s LZ}^{(0)}(1-Q_{\s LZ}^{(1)})}\left( e^{-i \chi_{\s 1}} + e^{-i \chi_{\s 3}}\right)\\
+i(1-Q_{\s LZ}^{(1)})\sqrt{1-Q_{\s LZ}^{(0)}}e^{-i(\chi_{\s 3} - \chi_{\s 2} + \chi_{\s 1})}.
\end{multline}
Different contributions to $A_{\s \uparrow \rightarrow \downarrow}$ describe partial
amplitudes to change the level at one transition and to not change the level at other
two transitions.  The survival probability is given by $Q_{\s LZ}=1-|A_{\s \uparrow \rightarrow \downarrow}|^2$.

If we assume that the phases $\chi_1$, $\chi_2$, and $\chi_3$  are completely uncorrelated the averaging over these phases yields
\begin{multline}
\label{pclassical}
Q_{\s LZ}^{(c)}=\left[Q_{\s LZ}^{(1)}\right]^2 Q_{\s LZ}^{(0)}+Q_{\s LZ}^{(0)}\left[1-Q_{\s LZ}^{(1)}\right]^2\\
+2Q_{\s LZ}^{(1)}\left[1-Q_{\s LZ}^{(1)}\right]\left[1-Q_{\s LZ}^{(0)}\right],
\end{multline}
which is nothing but the result Eq. (\ref{Qlzincoherent}) in which only the terms
$k=0$ and $k=\pm 1$ are kept. Now we take into account that the transitions $k=-1$ and
$k=1$ are identical, set $\chi_1=\chi_3$, and perform the averaging over the two phases.
This gives

\begin{equation}
\label{pclassical1}
Q_{\s LZ}=Q_{\s LZ}^{(c)}-2Q_{\s LZ}^{(1)}Q_{\s LZ}^{(0)}\left[1-Q_{\s LZ}^{(1)}\right].
\end{equation}
We note that the interference contribution to Eq. (\ref{pclassical1})
is negative. In fact, it
leads to a maximum in $Q_{\s LZ}(G)$ behavior even for the bare $Q_{\s LZ}=e^{-2}$, as
illustrated in Fig. \ref{classical}. Thus, it is the result Eq. (\ref{pclassical1}) that should be compared to the numerical calculations.
The results of these calculations are shown in Fig. \ref{bigomega},
where probability, $Q_{\s LZ}$,  calculated by solving
the system  Eq. (\ref{system1}) and averaging over $\kappa$
is shown for three oscillator frequencies versus the dimensionless coupling amplitude.
In the domain $\frac{G}{\omega}<0.1$ all three curves coincide and agree with the ``incoherent" and ``coherent" theoretical results shown in
Fig. \ref{classical}. The position of the maxima also agrees with the prediction of the ``coherent" theory Eq. (\ref{pclassical1}).
However, the peak value $Q_{\s LZ}\approx 0.6$ is higher than
 $Q_{\s LZ}=0.35$ predicted by the theory. The possible origin
 of the discrepancy is that the theory  Eq. (\ref{pclassical1}) takes into account only $k=0$ and $k=\pm 1$ intermediate transitions.
 Our overall conclusion is that non-monotonic behavior of $Q_{\s LZ}(G)$
 is the result of the interference of intermediate LZ transitions.


\section{Discussion and concluding remarks}

In the present paper we focused on the question: how the longitudinal coupling to a
harmonic oscillator affects the survival probability of the Landau-Zener
transition in a driven two-level system. 

On general grounds, one would expect the following answer to this question.
Weak coupling, by making the transition less adiabatic, increases the survival 
probability. At very strong coupling this probability should approach $1/2$, since
the memory about the initial state of the two-level system gets erased due to coupling. 
There is, however, the evidence that these expectations are not entirely correct.
Firstly, the exact result obtained in
Ref. \onlinecite{Wubs2006++} states that, for purely longitudinal coupling,
the effect is identically zero for {\em any} coupling strength.
Secondly, the numerical simulations of Ref. \onlinecite{Ashhab1},
which pertain to longitudinal coupling suggest that, at finite temperature, $Q_{\s LZ}(G)$ is a non-monotonic function with a maximum. 
The domain of parameters investigated 
in Ref. \onlinecite{Ashhab1} is intermediate in
all respects: the temperature, the oscillator frequency,
and the LZ tunneling gap were of the same order. In this
regime it is difficult to infer the underlying origin
of this maximum.

To establish the above physical picture,
we have adopted a strong assumption that the oscillator 
is highly excited, so that the
change of the excitation level in the course 
of the transition is relatively small.
Under this assumption, we studied the effect of coupling
on the LZ transition in two limiting cases of slow and 
fast oscillator. For the slow oscillator our analytical results for the survival
 probability are given by Eqs. (\ref{approximate}), (\ref{effective}), and Eq. (\ref{approximate1}).
For the fast oscillator they are given by Eq. (\ref{Qlzincoherent})and Eq. (\ref{pclassical1}).
We can now quantify the validity of our main assumption.
For both, the slow and the fast oscillator, the LZ transition
in the presence of coupling transforms into a  sequence of individual transitions.\cite{Garanin2013} This is illustrated in Fig. \ref{action}b for the slow
oscillator and in Fig. \ref{timeplotbessel} for the fast oscillator. 
The number of transitions for the slow oscillator is $N_s=4G\omega/\pi v$.
For the fast oscillator this number can be estimated by
equating the argument of the Bessel function in Eq. (\ref{Reduced2}) to
the index, i.e. $N_f=4G/\omega$.
Since each individual transition is associated with different level of the
oscillator, the criterion that the oscillator is highly excited can be quantified
as follows: the initial excitation level $n_0$ should be bigger than $N_s$ for the
slow oscillator, while for the fast oscillator it should be bigger than $N_f$.  

Although our analytical results essentially confirm the general expectations,
we find that, in both limits, $Q_{\s LZ}(G)$ approaches $1/2$ with oscillations.
These oscillations are the consequence of the interference of the amplitudes
corresponding to different pathways through multiple LZ transitions. 
It is likely that non-monotonic $Q_{\s LZ}(G)$ established in Ref. \onlinecite{Ashhab1}
is the consequence of this interference.   

Note finally, that throughout the paper we assumed the bare survival probability is small,
so that, unlike Ref. \onlinecite{GaraninPerturbative}, the perturbative treatment does not
apply.



%
%

\centerline{\bf Acknowledgements}
The work was supported by the Department of
Energy, Office of Basic Energy Sciences, Grant No. DE-
FG02-06ER46313.

\end{document}